\documentstyle[aps,prl,epsfig]{revtex}

\begin{document}
\title{Three-dimensional theory for interaction between atomic ensembles and
free-space light}
\author{L.-M. Duan$^{1,2}$\thanks{
Email: lmduan@caltech.edu}, J. I. Cirac$^{3}$, and P. Zoller$^{4}$}
\address{$^{1}$Institute for quantum information, mc 107-81, California
Institute of Technology, Pasadena, CA 91125-8100\\
$^{2}$Laboratory of quantum information, USTC, Hefei 230026, China\\
$^{3}$Max Planck Institut fuer Quantenoptik, Hans Kopfermannstr.
1, D-85748 Garching, Germany\\
$^{4}$Institut fuer Theoretische Physik, Universitaet Innsbruck,
A-6020 Innsbruck, Austria}
\maketitle

\begin{abstract}
Atomic ensembles have shown to be a promising candidate for implementations
of quantum information processing by many recently-discovered schemes. All
these schemes are based on the interaction between optical beams and atomic
ensembles. For description of these interactions, one assumed either a
cavity-QED model or a one-dimensional light propagation model, which is
still inadequate for a full prediction and understanding of most of the
current experimental efforts which are actually taken in the
three-dimensional free space. Here, we propose a perturbative theory to
describe the three-dimensional effects in interaction between atomic
ensembles and free-space light with a level configuration important for
several applications. The calculations reveal some significant effects which
are not known before from the other approaches, such as the inherent
mode-mismatching noise and the optimal mode-matching conditions. The
three-dimensional theory confirms the collective enhancement of the
signal-to-noise ratio which is believed to be one of the main advantage of
the ensemble-based quantum information processing schemes, however, it also
shows that this enhancement need to be understood in a more subtle way with
an appropriate mode matching method.
\end{abstract}

\section{Introduction}

Recently, many interesting schemes have been proposed which use atomic
ensembles with a large number of identical atoms as the basic system for
quantum state engineering and for quantum information processing. For
instance, one can use atomic ensembles for generation of substantial spin
squeezing \cite{1,2,3} and continuous variable entanglement \cite{4,5,6},
for storage of quantum light \cite{7,8,9,10,11}, for realization of scalable
long-distance quantum communication \cite{12}, and for efficient preparation
of many-party entanglement \cite{13}. The experimental candidates of atomic
ensembles can be either some cold or ultracold atoms in a trap \cite{2,10},
or a cloud of room-temperature atomic gas contained in a glass cell with
coated walls \cite{3,6,11}. The schemes based on atomic ensembles have some
special advantages compared with the quantum information schemes based on
the control of single \ particles: firstly, laser manipulation of atomic
ensembles without separate addressing of individual atoms is normally easier
than the coherent control of single particles; Secondly, and more
importantly, atomic ensembles with suitable level configurations could have
some kinds of collectively enhanced coupling to certain optical mode (called
the signal light mode) due to the many-atom interference effects. Thanks to
this enhanced coupling, we can obtain collective enhancement of the
signal-to-noise ratio (that is, the ratio between the controllable coherent
interaction and the uncontrollable noisy interactions) compared with the
single-particle case if we choose to manipulate the appropriate atomic and
optical modes. This collective enhancement plays an important role in all
the recent schemes based on the atomic ensembles.

To describe the interaction between atomic ensembles and optical beams, in
particular, to understand the collective enhancement of the signal-to-noise
ratio, normally one assumed a simple cavity-QED model \cite{7,8,12,1} or a
one-dimensional light propagation model \cite{9,4} with an independent
spontaneous emission rate for each atom. In contrast to this, most of the
current experiments are done with free-space atomic ensembles coupling
directly to the three-dimensional optical beams \cite{3,6,10,11}. It is not
obvious that the predictions from the simple models will always be valid for
this real much more complicate experimental situations. To fully predict and
understand the real experiments, one needs to answer various questions
associated with the three-dimensional interaction effects, for instance,
what is the inherent mode structure of the signal light when we have a
definite geometry of the atomic ensemble? Can we achieve a good mode
matching with the matching efficiency in principle arbitrarily close to one?
What is the noise magnitude associated with density fluctuation (induced by
the random initial distribution of the atom positions and the random atomic
motion) of the atomic ensemble? In the three-dimensional configuration, is
there still the collective enhancement of the signal-to-noise ratio?

In this paper, we will try to provide answers to the above important
questions for a Raman type $\Lambda $-level configuration which is useful
for scalable long-distance quantum communication \cite{12} and for
many-party entanglement generation \cite{13}. In general, it is very
challenging to build a full quantum theory from the first principle to
describe the interaction between the many-atom ensemble and the
infinite-mode optical field and to give definite answers to the above-listed
questions. In this paper, we will use a perturbative approach by assuming
that the Raman pumping laser is very weak. There are several motivations to
use a perturbative approach: firstly, the schemes in \cite{12,13} for
scalable quantum communication and for many-party entanglement generation
can work in the weak-pumping limit, so the perturbative approach describes a
useful realistic situation. Secondly, the perturbative calculations allow us
to investigate the three-dimensional effects and to give definite answers to
the questions listed above from the first principle without any doubtful
approximation. Finally, we expect the three-dimensional effects should be at
least to some extent independent of the power of the pumping laser, so the
perturbative calculations could give at least some indications for the
three-dimensional interaction picture in other regions.

The calculations reveal some significant results which are unexpected from
the simple models, for instance, it turns out that due to the density
fluctuation of the atomic ensemble, there will be two sources of noise for
the light-atomic-ensemble interaction: one is the spontaneous emission loss
and the other is the inherent mode mismatching noise (here, by ``inherent'',
we mean that this mode mismatching is not from any technical imperfection).
Thus, we need to use two quantities to describe the signal-to-noise ratios,
and need to keep a balance between these two sources of noise by appropriate
mode matching methods to optimize the setup for applications. The intuitive
mode matching method will result in a quite large inherent mode mismatching
noise. It is better to use some other mode matching methods which reduce the
mode matching noise at the price of increasing the spontaneous emission
loss. This can optimize the setup for some applications, such as the ones in
Refs. \cite{12,13}, since there the spontaneous emission loss has a far less
important influence. The calculations in this paper also demonstrate that in
the realistic three-dimensional configuration, one can still obtain large
collective enhancement of the signal-to-noise ratio for atomic ensembles
compared with the single-atom case, which is an important feature of this
kind of systems.

It is also helpful to make a comparison between the collective enhancement
of the signal-to-noise ratio we consider here and some other collective
optical effects, such as the superradiance. The similarity lies in that both
phenomena involve many-atom interference. However, there are also important
differences. For superradiance, it is a stimulated emission effect, and the
enhancement is in the emission speed. At the initial stage, coherence is
built between different atoms from the spontaneous emissions. After the
coherence has been built, superradiance can be understood even from a
classical interference picture. For the light-atomic-ensemble interaction
considered here, we are still completely in the spontaneous emission region
which needs a full \ quantum description, and the enhancement is not in the
emission speed, but in the signal-to-noise ratio when we only manipulate and
measure a definite atomic mode (called the symmetric collective atomic mode)
and a definite signal light mode (which is the optical mode collinear with
the Raman pumping light). The coupling between the above atomic and the
optical modes are coherent for different atoms (which means for this
coupling there is a certain phase relation between different atoms), while
the coupling of these two modes to other atomic and other optical modes
(called the noise modes) are inherent with random phase relations between
different atoms (the randomness comes from the random atom positions). We
have collective enhancement of the signal-to-noise ratio since the coherent
coupling interferes constructively for different atoms. Of course, in the
three-dimensional free-space, there are infinite optical modes, and the
signal light mode changes continuously to other noisy optical modes when we
continuously vary the solid angle. Due to the continuous change from the
coherent coupling to the inherent coupling, we have non-trivial mode
matching problem, and we get some inherent mode mismatching noise.

We also would like to mention that there are some early important works on
the transverse effects of the Raman interaction by using the
three-dimensional light propagation equations \cite{14}. However, it is not
clear to us how to use this approach to describe the collective enhancement
of the signal-to-noise ratio, and in particular, how to give definite
answers to the questions listed above. There is also some unpublished effort
to try to figure out the three-dimensional effects in another light-atom
interaction configuration \cite{15}. The interaction picture is not yet
clear to us.

This paper is arranged as follows: in Sec. II, we explain the interaction
scheme and the basic ideas of the applications of this interaction scheme
for quantum information processing. With the applications in mind, we can
focus our efforts on the most relevant quantities that we need to calculate.
We will also explain in this section the basic interaction picture between
the light and the atomic ensemble from our calculations. Then, in the next
section, we will describe the theoretical model for this light-atom
interaction from the first principle, and solve this model by using a
perturbative approach. In Sec. IV, we will discuss the properties of this
solution, in particular, we will define and calculate the mode structure of
the signal light, the spontaneous emission inefficiency, and the inherent
mode mismatching noise. Sec. V is devoted to the discussion of the
appropriate mode matching methods for real experiments. We show that by
appropriate mode matching methods, we can keep a balance between the two
sources of noise mentioned above in order to optimize this setup for
applications. Finally, in Sec. VI we summarize the results.

\section{The interaction scheme and its applications}

The basic element of our system is an atomic ensemble, which consists of a
cloud of $N_{a}$ identical atoms with the relevant level structure shown in
Fig. 1. A pair of metastable lower states $|g\rangle $ and $|s\rangle $ can
correspond, for instance, to hyperfine or Zeeman sublevels of electronic
ground state of alkali atoms. The relevant coherence between the levels $%
|g\rangle $ and $|s\rangle $ can be maintained for a long time, as has been
demonstrated both experimentally \cite{6,10,11} and theoretically \cite%
{7,8,9}. All the atoms are initially prepared in the ground state $\left|
g\right\rangle $ through optical pumping.
\begin{figure}[tbp]
\epsfig{file=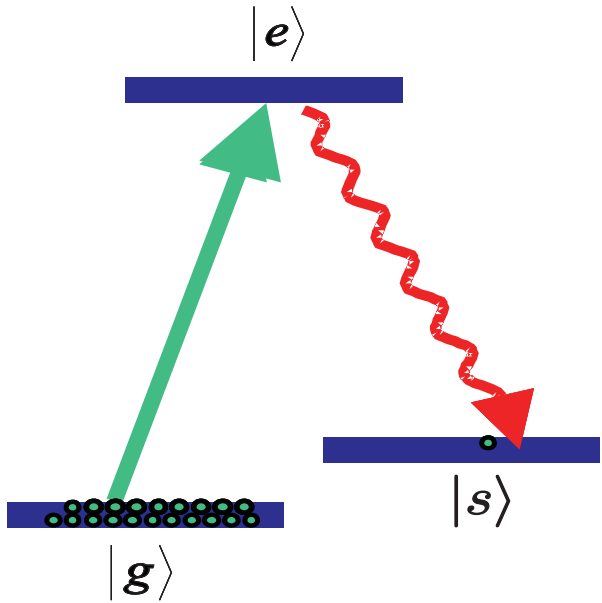,width=8cm}
\caption{The relevant atomic level structure with $\left| g\right\rangle $,$%
\left| s\right\rangle $, a pair of ground or metastable states, and $\left|
e\right\rangle $ , the excited state.}
\end{figure}

The ensemble is then illuminated by a weak pumping laser pulse which couples
the transition $|g\rangle \rightarrow |e\rangle $ with a large detuning $%
\Delta $, and we look at the spontaneous emission light from the transition $%
|e\rangle \rightarrow |s\rangle $, whose polarization and/or frequency are
assumed to be different to the pumping laser. There are two interaction
configurations: in the first configuration, the pumping laser is shined on
all the atoms so that each atom has an equal small probability to be excited
into the state $\left| s\right\rangle $ through the Raman transition. In the
second configuration, as is reported in the experiment \cite{6}, the pumping
laser is focused with a transverse area smaller than the transverse area of
the atomic ensemble, so only part of the atoms are illuminated by the laser
at each instant. However, during the light-atom interaction period, all the
atoms in the ensemble are moving fast, and they frequently enter and leave
the interaction region. As a result, each atom still has an equal small
probability to be excited into the state $\left| s\right\rangle $.

As we will show in the next section, after the atomic gas interacts with a
weak pumping laser, there will be a special atomic mode $s_{s}$ called the
symmetric collective atomic mode, and a special optical spontaneous emission
mode $a_{s}$ (which couples to the transition $|e\rangle \rightarrow
|s\rangle $) called the signal light mode. The symmetric collective atomic
mode $s_{s}$ and the signal light mode $a_{s}$ are defined respectively by
\begin{equation}
s_{s}\equiv \left( 1/\sqrt{N_{a}}\right) \sum_{i=1}^{N_{a}}\left|
g\right\rangle _{i}\left\langle s\right| ,  \label{1}
\end{equation}%
\begin{equation}
a_{s}=\int f_{{\bf k}}^{\ast }a_{{\bf k}}d^{3}{\bf k,}  \label{2}
\end{equation}%
where $a_{{\bf k}}$ represents the the plane wave mode with the wave vector $%
{\bf k}$ (we have used the plane wave modes as the eigenmodes for the
expansion of the optical spontaneous emission field). The operators $a_{{\bf %
k}}$ satisfy the standard commutation relations $\left[ a_{{\bf k}},a_{{\bf k%
}^{\prime }}^{\dagger }\right] =\delta \left( {\bf k-k}^{\prime }\right) $,
and $f_{{\bf k}}^{\ast }$ is the normalized signal mode function whose
explicit form will be specified in Sec. IV. We just need to mention here
that $a_{s}$ represents the spontaneous emission light which is basically
collinear with the pumping laser, with a distribution only over a very small
solid angle. The particularity of the modes $s_{s}$ and $a_{s}$ comes from
the fact that they are dominantly correlated with each other, which means,
if a atom is excited to the symmetric collective mode $s_{s}$, the
accompanying spontaneous emission photon will most probably go to the signal
light mode $a_{s}$, and vice versa. There are still many other atomic modes
in the ensemble and infinite other optical modes in the spontaneous emission
field, and these atomic and optical modes are correlated with each other in
a complicated way. However, the correlation between the good modes $s_{s}$
and $a_{s}$ is quite ``pure'', and these two modes are only weakly
correlated with the other atomic and optical modes which contribute to
noise. The correlation between the atomic mode $s_{s}$ and other optical
modes contributes to the spontaneous emission loss, and the correlation
between the signal mode $a_{s}$ and other atomic modes contributes to the
inherent mode mismatching noise. The interaction picture for this system is
schematically shown by Fig. 2.
\begin{figure}[tbp]
\epsfig{file=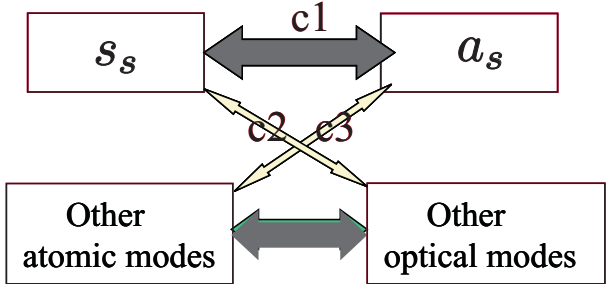,width=8cm} \caption{For a many-atom
ensemble, The intuitive interaction picture between the atomic
modes and the free-space optical modes, where c1 represents the
''good'' correlation between the collective atomic mode and the
signal light mode, c2 represents the spontaneous emission loss,
and c3 represents the inherent mode mismatching noise. Broader
connections stand for stronger correlations between the
excitations in the corresponding modes. }
\end{figure}

The applications of this system for quantum information processing exactly
comes from the almost pure correlation between the modes $s_{s}$ and $a_{s}$%
. If we neglect the weak correlations between the good modes $s_{s},a_{s}$
and the other noisy atomic and optical modes, that is, we neglect the two
sources of noise illustrated in Fig. 2, all the noisy modes can be traced
over, and we get effectively a two-mode problem. The excitations in the
modes $s_{s}$ and $a_{s}$ can be both separately measured. For the signal
mode $a_{s}$, this is done through a single-photon detector with an
appropriate mode-matching; and for the atomic mode $s_{s}$, this can be done
by first mapping the atomic excitation to an excitation in the signal mode
through a repumping laser pulse \cite{12}, and then detecting it again
through a single-photon detector. Using the almost pure correlation between
the modes $s_{s}$ and $a_{s}$, one can generate some preliminarily
entanglement between two distant atomic ensembles by only linear optics
means, which forms the important first step for applications of this setup
for different kinds of quantum information processing tasks detailed in
Refs. \cite{12,13}. Here, we will briefly explain the basic ideas on how to
entangle atomic ensembles using the correlation between the modes $s_{s}$
and $a_{s}$. This explanation helps us to define the most relevant
quantities for the applications that we need to calculate.

The setup for entanglement generation between the two distant atomic
ensembles L and R is shown by Fig. 3. We apply simultaneously two short
Raman pumping pulses on the ensembles L and R, respectively, so that for
each \ ensemble the light scattered to the signal mode $a_{s}$ has a mean
photon number much smaller than $1$. The signal modes are then coupled to
optical channels (such as fibers) through mode matching after the filters,
which are polarization and frequency selective to filter the pumping light.
The signal pulses after the transmission channels interfere at a 50\%-50\%
beam splitter BS, with the outputs detected respectively by two
single-photon detectors D1 and D2. If either D1 {\it or} D2 registers one
photon, the process is finished. Otherwise, we first apply a repumping laser
pulse to the transition $\left| s\right\rangle \rightarrow \left|
e\right\rangle $ on the ensembles L and R to set the atoms back to the
ground state $\left| g\right\rangle $, then the same Raman driving pulses as
the first round are applied to the transition $\left| g\right\rangle
\rightarrow \left| e\right\rangle $ and we detect again the photon number of
the signal modes after the beam splitter. This process is repeated until
finally we have a click in either D1 or D2 detector.
\begin{figure}[tbp]
\epsfig{file=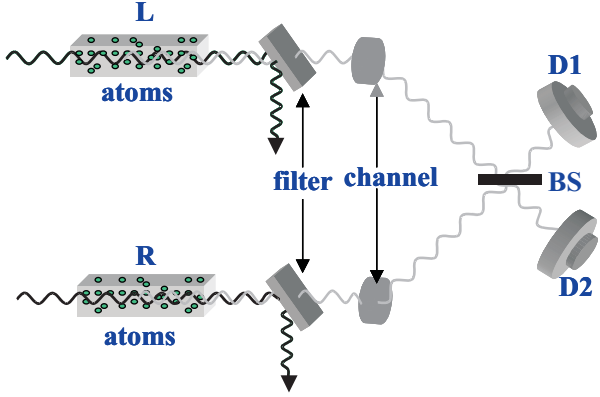,width=8cm} \caption{The schematic setup for
generating entanglement between two distant atomic ensembles L and
R.}
\end{figure}

We show now how the entanglement is generated between the ensembles L and R
if we successfully register one photon in D1 or D2 detector. To understand
this, let us first look at one ensemble. Before the Raman pumping pulse, the
collective atomic mode $s_{s}$ and the signal light mode $a_{s}$ are in
vacuum states, which are denoted respectively\ by $\left| 0_{a}\right\rangle
$, $\left| 0_{p}\right\rangle $. If we neglect the small noisy correlations,
after the weak pumping, the state of the the modes $s_{s}$ and $a_{s}$ has
the following form (see the next section for the detailed derivation)

\begin{equation}
\left| \phi \right\rangle =\left| 0_{a}\right\rangle \left|
0_{p}\right\rangle +\sqrt{p_{c}}s_{s}^{\dagger }a_{s}^{\dagger }\left|
0_{a}\right\rangle \left| 0_{p}\right\rangle +o\left( p_{c}\right) ,
\label{3}
\end{equation}%
where $p_{c}\ll 1$ is the probability for one atom excited to the collective
atomic mode $s_{s}$, and $o\left( p_{c}\right) $ represents the high-order
terms with more excitations whose probabilities are equal to or smaller than
$p_{c}^{2}$.

In Fig. 3, the pumping pulses excite both ensembles simultaneously, the
whole system is thus described by the state $\left| \phi \right\rangle
_{L}\otimes \left| \phi \right\rangle _{R}$, where $\left| \phi
\right\rangle _{L}$ and $\left| \phi \right\rangle _{R}$ are given by Eq.
(1) with all the operators and states distinguished by the subscript $L$ or $%
R$, respectively. The two signal light modes are superposed at the beam
splitter, and a photodetector click in either D1 or D2 measures respectively
the operators $a_{+}^{\dagger }a_{+}$ or $a_{-}^{\dagger }a_{-}$ with $%
a_{\pm }=\left( a_{sL}\pm e^{i\varphi }a_{sR}\right) /\sqrt{2}$. Here, $%
\varphi $ denotes the difference of the phase shifts in the two-side optical
channels. Conditional on the detector click, we should apply a projection
operator $a_{+}$ or $a_{-}$ onto the whole state $\left| \phi \right\rangle
_{L}\otimes \left| \phi \right\rangle _{R}$. If we neglect the high-order
corrections in Eq. (3), the projected state of the ensembles L and R thus
has the form
\begin{equation}
\left| \Psi _{\varphi }\right\rangle _{LR}^{\pm }=\left( s_{sL}^{\dagger
}\pm e^{i\varphi }s_{sR}^{\dagger }\right) /\sqrt{2}\left|
0_{a}\right\rangle _{L}\left| 0_{a}\right\rangle _{R},  \label{4}
\end{equation}%
which is maximally entangled in the excitation number basis. The generation
of this kind of state forms the basis of further applications in quantum
information processing \cite{12,13}. If we take into account the high-order
terms in Eq. (3), the fidelity between the actually generated state and the
ideal state (4) will decrease by an amount proportional to $p_{c}$, and this
decrease will contribute to the final fidelity imperfection of the schemes
in Refs. \cite{12,13}. For the applications described by these schemes, we
need to fix the fidelity imperfection to be small, which means that we
should keep a small excitation probability $p_{c}$ by controlling the
intensity of the Raman pumping laser.

Now we look at the influence of the two types of noisy correlations
illustrated in Fig. 2. The spontaneous emission loss can be quantified by
the probability $p_{\text{spon}}$ of the event that the scattered photon
goes to some other optical modes instead of the signal mode $a_{s}$ while
the accompanying atom is excited to the collective atomic mode $s_{s}$.
Without the spontaneous emission loss, for each round of Raman pumping, we
succeed with a probability $2p_{c}$ to get a detector click, which prepares
the entangled state (4). However, the spontaneous emission loss means that
even if an atom is excited to the mode $s_{s}$ (which has a probability $%
p_{c}$ for each ensemble), the accompanying photon, with a probability $1-p_{%
\text{spon}}$, cannot be registered. For entanglement generation, the
fidelity imperfection to the ideal state (4) is given by the excitation
probability of the atomic mode $s_{s}$, which is now $p_{c}/\left( 1-p_{%
\text{spon}}\right) $ instead of $p_{c}$ (we always use $p_{c}$ to denote
the possibility of the good event described by Eq. (3)). To fix the fidelity
imperfection to be small, in the presence of the spontaneous emission loss,
we need to further decrease the excitation probability $p_{c}$ by a factor
of $1-p_{\text{spon}}$, which means that we have a smaller probability to
register one signal photon. So, as a result of this noise, the preparation
efficiency (the success probability of the scheme) is decreased by a factor
of $1-p_{\text{spon}}$ when we fix the fidelity imperfection.

The inherent mode-mismatching noise can be quantified by the probability $p_{%
\text{mode}}$ of the event that an atom is excited to some other atomic
modes instead of the collective mode $s_{s}$ while the accompanying photon
goes to the right signal mode $a_{s}$. In this case, we can register a
photon from the single-photon detectors, but the atomic mode $s_{s}$ will be
actually still in the initial vacuum state $\left| 0_{a}\right\rangle $. So,
this noise will add some vacuum component to the generated state between the
ensembles L and R. In the presence of this noise, the generated state
between the ensembles L and R is mixed with the form
\begin{equation}
\rho _{LR}\left( c_{0},\varphi \right) =\frac{1}{c_{0}+1}\left( c_{0}\left|
0_{a}0_{a}\right\rangle _{LR}\left\langle 0_{a}0_{a}\right| +\left| \Psi
_{\varphi }\right\rangle _{LR}^{\text{ }+}\left\langle \Psi _{\varphi
}\right| \right) ,  \label{5}
\end{equation}%
where the vacuum coefficient $c_{0}$ is basically given by the conditional
probability $p_{\text{mode}}$ for this noise contribution. Actually, as has
been shown in \cite{12}, there are other sources of noise which can
contribute to the vacuum coefficient $c_{0}$, such as the detector dark
counts, and the detector inefficiency in the succeeding entanglement
connection scheme detailed in Ref. \cite{12}. It has also been shown there
the vacuum component noise will be finally automatically purified, and thus
has no influence on the final communication fidelity. However, it has a more
significant influence on the efficiency than the spontaneous emission loss.
For applications in Refs. \cite{12,13}, to get a better overall efficiency,
it is better to keep a balance between the mode mismatching noise $p_{\text{%
mode}}$ and the spontaneous emission inefficiency $p_{\text{spon}}$, with $%
p_{\text{mode}}$ significantly smaller than $p_{\text{spon}}$.

It is helpful to make a comparison here with the single-atom case. For a
single atom interacting with the free-space light with the same level
configuration as shown in Fig. 1, there is only one atomic mode given by $%
s=\left| g\right\rangle \left\langle s\right| $, but there are still
infinite optical modes. We can still identify the forward scattered optical
mode $a_{s}$ as the signal mode. The interaction picture is then shown
intuitively by Fig. 4. For a single atom, there is no inherent
mode-mismatching noise, but the spontaneous emission inefficiency becomes
much larger. If the atom is excited to the level $\left| s\right\rangle $
(that is, to the mode $s$) through the Raman laser pumping, the accompanying
photon has a probability to go to all the possible directions \cite{16}, and
thus has only a very small possibility to go to the signal mode $a_{s}$. We
will calculate in the following sections the spontaneous emission
inefficiencies for the signal atom case as well as for the atomic ensemble
case, and compare them with the different mode-matching methods to see
whether there is collective enhancement of the signal-to-noise ratio for the
many-atom ensemble.
\begin{figure}[tbp]
\epsfig{file=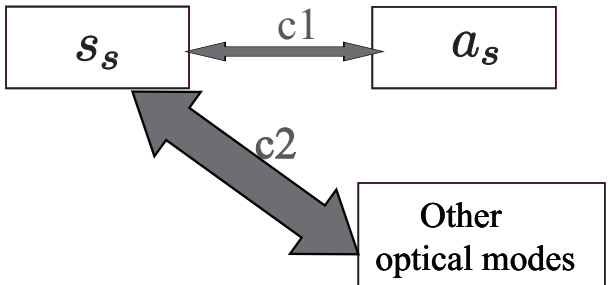,width=8cm} \caption{The intuitive
interaction picture for the single-atom case with the
corresponding symbols having the same meaning as Fig. 2. There is
only one atomic mode, but the spontaneous emission loss
represented by c2 becomes much larger compared with the good
correlation c1. }
\end{figure}

\section{The theoretical model for the light-atomic-ensemble interaction and
its solution}

We will now go to the detailed description of the light-atomic-ensemble
interaction. From this description, we will calculate the signal-light mode
structure $f_{{\bf k}}$, the spontaneous emission inefficiency $p_{\text{spon%
}}$, and the inherent mode-mismatching noise $p_{\text{mode}}$. In the
interaction configuration shown by Fig. 1, the Raman pumping laser can be
described classically by neglecting its small quantum variance, and is
assumed to be propagating basically along the $z$ direction. So, we can
write its amplitude as $\varepsilon _{ge}\left( {\bf r},t\right) =u\left(
{\bf r},t\right) e^{i\left( k_{0}z-\omega _{0}t\right) }$, where $\omega
_{0}=k_{0}c=2\pi c/\lambda _{0}$ is the carrier frequency and $u\left( {\bf r%
},t\right) $\ is a slow varying function of the coordinate ${\bf r}$ and the
time $t$. The field $\varepsilon _{se}$ coupling to the transition $%
|e\rangle \rightarrow |s\rangle $ should be described quantum mechanically,
and we expand it into plane wave modes $\varepsilon _{se}\left( {\bf r}%
,t\right) \propto \int a_{{\bf k}}e^{i\left( {\bf k\cdot r}-\omega _{{\bf k}%
}t\right) }d^{3}{\bf k}$ with $\omega _{{\bf k}}=\left| {\bf k}\right| c$,
and $a_{{\bf k}}$, the corresponding annihilation operator. Then, the
Hamiltonian describing the light-atom interaction has the following form in
the interaction picture (setting $\hbar =1$)%
\begin{eqnarray}
H\left( t\right) &=&\Delta \sum_{i=1}^{N_{a}}\sigma _{ee}^{i}+\left\{
g_{eg}\sum_{i=1}^{N_{a}}\sigma _{eg}^{i}u\left( {\bf r}_{i},t\right)
e^{ik_{0}z}\right.  \label{e6} \\
&&\left. +\sum_{i=1}^{N_{a}}\sigma _{se}^{i}\int g_{se}^{{\bf k}}a_{{\bf k}%
}^{\dagger }e^{-i\left[ {\bf k\cdot r}_{i}-\left( \omega _{{\bf k}}-\omega
_{0}+\omega _{sg}\right) t\right] }d^{3}{\bf k+H.c.}\right\} ,  \nonumber
\end{eqnarray}%
where the detuning $\Delta =\omega _{eg}-\omega _{0}$ with $\omega
_{eg}=\omega _{e}-\omega _{g}$, the frequency difference between the atomic
levels, $\sigma _{\mu \nu }^{i}=\left| \mu \right\rangle _{i}\left\langle
\nu \right| $\ ($\mu ,\nu =g,e,s$) are the transition operators of the $i$th
atom, ${\bf r}_{i}$ is the coordinate of the $i$th atom, and $g_{eg},g_{se}^{%
{\bf k}}$ are the coupling coefficients which are proportional to the dipole
moments of the corresponding transitions. The coefficient $g_{se}^{{\bf k}}$
depends in general on the direction of the wave vector ${\bf k}$ by the
dipole pattern. In Eq. (6), the carrier frequency of the spontaneous
emission field $\varepsilon _{se}\left( {\bf r},t\right) $ is given by $%
\omega _{0}-\omega _{sg}$, and its relevant frequency width can be estimated
by the natural width $\Gamma $ of the excited level $|e\rangle $, which
means that the modes $a_{{\bf k}}$ with the frequency difference between $%
\omega _{{\bf k}}$ and $\omega _{0}-\omega _{sg}$ much larger than the width
$\Gamma $ will have negligible influence on the system dynamics. We neglect
in Eq. (6) the spontaneous emission back to the level $|g\rangle $ since it
is not important for our purpose (with the detection method specified in
Sec. II), and has no influence on all of our results.

If both the natural width $\Gamma $ and the frequency spreading of the
pumping laser $u\left( {\bf r},t\right) $ are significantly smaller than the
detuning $\Delta $, we can adiabatically eliminate the upper level $%
|e\rangle $ through the standard technique, and the resulting adiabatic
Hamiltonian is given by%
\begin{eqnarray}
H\left( t\right) &=&-\left\{ \sum_{i=1}^{N_{a}}\sigma _{sg}^{i}u\left( {\bf r%
}_{i},t\right) \int \frac{g_{eg}g_{se}^{{\bf k}}}{\Delta }a_{{\bf k}%
}^{\dagger }e^{-i\left[ \Delta {\bf k\cdot r}_{i}-\Delta \omega _{{\bf k}}t%
\right] }d^{3}{\bf k}\right.  \nonumber \\
&&\left. {\bf +H.c.}\right\} -\frac{\left| g_{eg}\right| ^{2}}{\Delta }%
\sum_{i=1}^{N_{a}}\sigma _{gg}^{i}\left| u\left( {\bf r}_{i},t\right)
\right| ^{2},  \label{7}
\end{eqnarray}%
where $\Delta \omega _{{\bf k}}=\omega _{{\bf k}}-\left( \omega _{0}-\omega
_{sg}\right) $ and $\Delta {\bf k=k-}k_{0}{\bf z}_{0}$ with ${\bf z}_{0}$,
the unit vector in the $z$ direction. We have neglected in Eq. (7) the Stark
shift of the level $|s\rangle $ since it is much smaller than the other
terms (The spontaneous emission field is much weaker than the pumping
field). The last term in Eq. (7) (the Stark shift of the level $|g\rangle $%
)can be eliminated if we make a phase rotation of the basis $\left\{
|g\rangle ,|s\rangle \right\} $ which will transform $\sigma _{sg}^{i}$ to $%
\sigma _{sg}^{i}e^{i\left( \left| g_{eg}\right| ^{2}/\Delta \right)
\int_{0}^{t}\left| u\left( {\bf r}_{i},\tau \right) \right| ^{2}d\tau }$.
Thus, we can simply drop off the last term, and replace $u\left( {\bf r}%
_{i},t\right) $ by $u^{\prime }\left( {\bf r}_{i},t\right) =u\left( {\bf r}%
_{i},t\right) e^{i\left( \left| g_{eg}\right| ^{2}/\Delta \right)
\int_{0}^{t}\left| u\left( {\bf r}_{i},\tau \right) \right| ^{2}d\tau }$. In
the following, for simplicity of the symbol, we will denote $u^{\prime
}\left( {\bf r}_{i},t\right) $ still by $u\left( {\bf r}_{i},t\right) $, and
denote $-g_{eg}g_{se}^{{\bf k}}/\Delta $ by a single coefficient $g_{{\bf k}%
} $.

At the beginning, all the atoms and in the ground state $|g\rangle $, and
all the optical modes $a_{{\bf k}}$ are in the vacuum state. We denote this
initial state of all the atomic and optical modes by $|$vac$\rangle $. Then,
if the Raman pumping laser $u\left( {\bf r},t\right) $ is short and weak
enough, we can expand the final state $|\Psi _{f}\rangle $ into a
perturbative expansions. To the second order of the perturbation, the final
state has the form%
\begin{eqnarray}
|\Psi _{f}\rangle &=&\left\{ 1-i\int_{0}^{t_{0}}H\left( \tau \right) d\tau
\right.  \nonumber \\
&&\left. -\frac{1}{2}T\left[ \int_{0}^{t_{0}}\int_{0}^{t_{0}}H\left( \tau
_{1}\right) H\left( \tau _{2}\right) d\tau _{1}d\tau _{2}\right] \right\} |%
\text{vac}\rangle  \label{e8}
\end{eqnarray}%
where $t_{0}$ is the duration of the Raman pulse, and $T\left[ \cdots \right]
$ denotes the time-ordered product.

At this point, we should note that for an atomic vapor, the atom positions $%
{\bf r}_{i}$ are randomly distributed, and during the light-atom
interaction, the atoms are moving fast in the room-temperature \cite{17}.
Therefore, we should treat ${\bf r}_{i}$ in the Hamiltonian (7) as
stochastic variables. We make the following two assumptions about the
properties of these variables: (i) For different atoms $i$ and $j$, ${\bf r}%
_{i}$ and ${\bf r}_{i}$ do not correlate with each other; (ii) Different
stochastic variables ${\bf r}_{i}$ obey the same probability distribution
which is determined by the geometry of the atomic cell. Note that these two
assumptions are well satisfied by the room-temperature atomic gas. If the
atom positions ${\bf r}_{i}$ behave as classical random variables, we should
take an average of the final state $|\Psi _{f}\rangle $ over the joint
probability distribution of these variables, and the resulting state becomes
mixed with
\begin{equation}
\rho _{f}=\left\langle |\Psi _{f}\rangle \left\langle \Psi _{f}\right|
\right\rangle _{\left\{ {\bf r}_{i}\right\} },  \label{9}
\end{equation}%
where the symbol $\left\langle \cdots \right\rangle _{\left\{ {\bf r}%
_{i}\right\} }$ denotes the average over all the variables ${\bf r}_{i}$.

To take the average in Eq. (9), first we note that the Hamiltonian (7) can
be written as
\begin{equation}
H\left( t\right) =\sum_{i=1}^{N_{a}}H_{i}\left( {\bf r}_{i},t\right)
\label{10}
\end{equation}%
with
\begin{equation}
H_{i}\left( {\bf r}_{i},t\right) =\sigma _{sg}^{i}\int g_{{\bf k}}a_{{\bf k}%
}^{\dagger }e^{i\Delta \omega _{{\bf k}}t}\left\{ u\left( {\bf r}%
_{i},t\right) e^{-i\Delta {\bf k\cdot r}_{i}}\right\} d^{3}{\bf k.}
\label{11}
\end{equation}%
Only the term in the large bracket of the Hamiltonian $H_{i}$ depends on the
variable ${\bf r}_{i}$. By the property (i) of the variables ${\bf r}_{i}$,
we have%
\begin{eqnarray}
\left\langle H_{i}H_{j}\right\rangle _{\left\{ {\bf r}_{i}\right\} }
&=&\left\langle H_{i}\right\rangle _{\left\{ {\bf r}_{i}\right\}
}\left\langle H_{j}\right\rangle _{\left\{ {\bf r}_{i}\right\} }  \nonumber
\\
&&+\delta _{ij}\left\{ \left\langle H_{i}^{2}\right\rangle _{\left\{ {\bf r}%
_{i}\right\} }-\left\langle H_{i}\right\rangle _{\left\{ {\bf r}_{i}\right\}
}^{2}\right\} .  \label{12e}
\end{eqnarray}%
By the property (ii) of the variables ${\bf r}_{i}$, we know that $%
\left\langle u\left( {\bf r}_{i},t\right) e^{-i\Delta {\bf k\cdot r}%
_{i}}\right\rangle _{\left\{ {\bf r}_{i}\right\} }$\ becomes independent of
the atom index $i$, so the average of the Hamiltonian $H\left( t\right) $
has the simple form
\begin{eqnarray}
\left\langle H\left( t\right) \right\rangle _{\left\{ {\bf r}_{i}\right\} }
&=&\sqrt{N_{a}}s_{s}  \label{13} \\
&&\times \int g_{{\bf k}}a_{{\bf k}}^{\dagger }e^{i\Delta \omega _{{\bf k}%
}t}\left\langle u\left( {\bf r}_{i},t\right) e^{-i\Delta {\bf k\cdot r}%
_{i}}\right\rangle _{\left\{ {\bf r}_{i}\right\} }d^{3}{\bf k,}  \nonumber
\end{eqnarray}%
where $s_{s}$ is the symmetrical collective atomic operators defined in Eq.
(1). We will write the initial vacuum state $|$vac$\rangle $ as a tensor
product of the atomic part $|$vac$\rangle _{a}$ and the optical part $|$vac$%
\rangle _{p}$, and denote the atomic state $\sigma _{sg}^{i}|$vac$\rangle
_{a}$ simply by $|s\rangle _{i}$. With this notation, a combination of Eqs.
(8)-(13) yields the following form for the averaged state $\rho _{f}$
\begin{equation}
\rho _{f}=\left( 1-p_{2}-p_{c}\right) |\Psi _{\text{eff}}\rangle
\left\langle \Psi _{\text{eff}}\right| +p_{2}\rho _{n}+o\left( p_{2}\right) ,
\label{14}
\end{equation}%
where the effective pure state $|\Psi _{\text{eff}}\rangle $
(not-normalized) is%
\begin{equation}
|\Psi _{\text{eff}}\rangle =\left( 1-i\int_{0}^{t_{0}}\left\langle H\left(
\tau \right) \right\rangle _{\left\{ {\bf r}_{i}\right\} }d\tau \right)
|vac\rangle ,  \label{15}
\end{equation}%
with $\left\langle \Psi _{\text{eff}}\right| \Psi _{\text{eff}}\rangle
=1+p_{c}$, and the noise component $p_{2}\rho _{n}$ is given by%
\begin{eqnarray}
p_{2}\rho _{n} &=&\sum_{i=1}^{N_{a}}|s\rangle _{i}\left\langle s\right|
\otimes \int_{0}^{t_{0}}\int_{0}^{t_{0}}d\tau _{1}d\tau _{2}\int d^{3}{\bf k}%
\int d^{3}{\bf k}^{\prime }  \nonumber \\
&&\times g_{{\bf k}}g_{{\bf k}^{\prime }}^{\ast }e^{i\left( \Delta \omega _{%
{\bf k}}\tau _{1}-\Delta \omega _{{\bf k}^{\prime }}\tau _{2}\right) }a_{%
{\bf k}}^{\dagger }|\text{vac}\rangle _{p}\left\langle \text{vac}\right| a_{%
{\bf k}^{\prime }}  \label{16} \\
&&\times \left\{ \left\langle u\left( {\bf r}_{i},\tau _{1}\right) u^{\ast
}\left( {\bf r}_{i},\tau _{2}\right) e^{-i\left( {\bf k-k}^{\prime }\right)
{\bf \cdot r}_{i}}\right\rangle _{\left\{ {\bf r}_{i}\right\} }\right.
\nonumber \\
&&\left. -\left\langle u\left( {\bf r}_{i},\tau _{1}\right) e^{-i\Delta {\bf %
k\cdot r}_{i}}\right\rangle _{\left\{ {\bf r}_{i}\right\} }\left\langle
u\left( {\bf r}_{i},\tau _{2}\right) e^{i\Delta {\bf k}^{\prime }{\bf \cdot r%
}_{i}}\right\rangle _{\left\{ {\bf r}_{i}\right\} }\right\} ,  \nonumber
\end{eqnarray}%
with the value of $p_{2}$ determined by the normalization of $\rho _{n}$,
i.e., by $tr\left( \rho _{n}\right) =1$. In Eq. (14), $o\left( p_{2}\right) $
represents the higher-order terms compared with the magnitude of $p_{2}$ and
$p_{c}$, which will be neglected in the following. We also ignored the
second-order cross-terms proportional to $\sigma _{sg}^{i}a_{{\bf k}%
}^{\dagger }|$vac$\rangle \left\langle \text{vac}\right| $ or $|$vac$\rangle
\left\langle \text{vac}\right| \sigma _{gs}^{i}a_{{\bf k}}$ in writing Eq.
(14), since they have no contributions to the quantities we will calculate
in the below. The solution (14) serves as the starting point for discussions
of various properties of this system in the following section.

\section{Properties of the light-atomic-ensemble interaction}

Now, let us analyze the properties of the light-atomic-ensemble interaction
revealed by the solution (14). First, we look at the effective pure state $%
|\Psi _{\text{eff}}\rangle $ by neglecting the noise component $p_{2}\rho
_{n}$. The state $|\Psi _{\text{eff}}\rangle $ has exactly the same form as
the ideal state (3) which is the starting point for all the applications. To
see this clearly, we can write $|\Psi _{\text{eff}}\rangle $ as $|\Psi _{%
\text{eff}}\rangle =\left( 1+\sqrt{p_{c}}s_{s}^{\dagger }a_{s}^{\dagger
}\right) |$vac$\rangle $, by defining the signal mode%
\begin{eqnarray}
a_{s}^{\dagger } &\equiv &\int f_{{\bf k}}a_{{\bf k}}^{\dagger }d^{3}{\bf k}
\label{17} \\
&{\bf \equiv }&-i\frac{\sqrt{N_{a}}}{\sqrt{p_{c}}}\int d^{3}{\bf k}%
\int_{0}^{t_{0}}d\tau g_{{\bf k}}a_{{\bf k}}^{\dagger }e^{i\Delta \omega _{%
{\bf k}}\tau }\left\langle u\left( {\bf r}_{i},\tau \right) e^{-i\Delta {\bf %
k\cdot r}_{i}}\right\rangle _{\left\{ {\bf r}_{i}\right\} }{\bf ,}  \nonumber
\end{eqnarray}%
where the value of $p_{c}$ is determined by the normalization of the
operator $a_{s}^{\dagger }$, i.e., by the condition $\left[
a_{s},a_{s}^{\dagger }\right] =1$.

We are particularly interested in the spatial structure of the signal mode $%
a_{s}^{\dagger }$, since one needs to know this spatial structure to design
mode-matching in practical experiments. The integration $\int d^{3}{\bf k}$
can be expressed as $\int d^{3}{\bf k=}\int_{\omega _{0}-\omega _{sg}-\delta
/2}^{\omega _{0}-\omega _{sg}+\delta /2}d\omega \int_{4\pi }d\Omega $, where
$d\Omega =\sin \theta d\theta d\varphi $ represents the infinitesimal solid
angle and $\delta $ is the bandwidth of the spontaneous emission field which
is in the order of the natural width $\Gamma $ of the excited level $%
|e\rangle $. The spatial structure of the mode $a_{s}^{\dagger }$ is
determined by the superposition coefficient $f_{{\bf k}}=f\left( \omega
,\Omega \right) $ as a function of the solid angle $\Omega $. Typically, the
size $L$ of the atomic ensemble (centimeter long or less) satisfies the
condition $\Gamma L/c\ll 1$, and $\omega _{sg}L/c\ll 1$ ($\omega _{eg}$ is
either around GHz or zero, depending on one uses the hyperfine level or the
Zeeman sublevel for the $|s\rangle $ state). In this case, $\Delta {\bf %
k\cdot r}_{i}=\left( {\bf k-}k_{0}{\bf z}_{0}\right) {\bf \cdot r}_{i}\simeq
k_{0}\left( {\bf k/}\left| {\bf k}\right| -{\bf z}_{0}\right) {\bf \cdot r}%
_{i}$, which only depends on the direction of ${\bf k}$, and becomes
independent of the frequency $\omega $. We assume that the Raman pumping
field can be factorized as $u\left( {\bf r},t\right) =u_{\perp }\left( {\bf r%
}\right) \mu _{l}\left( z,t\right) $, where $\mu _{l}\left( z,t\right) $ is
a slowly changing function of the coordinate, and can be well approximated
as a constant along the size of the atomic ensemble. With this condition,
the superposition function $f\left( \omega ,\Omega \right) $ is factorized
as a product of the frequency part $f_{\omega }\left( \omega \right) $ and
the spatial part $f_{\Omega }\left( \Omega \right) $, and the spatial part $%
f_{\Omega }\left( \Omega \right) $ becomes independent of the emission
frequency $\omega $ and the interaction time $t_{0}$. To have a more
explicit expression of the spatial structure $f_{\Omega }\left( \Omega
\right) $ of the signal mode $a_{s}^{\dagger }$, we assume that the atoms
are distributed by the normalized distribution function $p_{\text{dis}%
}\left( {\bf r}\right) $ (with $\int p_{\text{dis}}\left( {\bf r}\right)
d^{3}{\bf r}=1$), which is determined by the geometry of the ensemble. As we
mentioned before, the interaction coefficient $g_{{\bf k}}$ normally also
depends on the direction of the wave vector ${\bf k}$ in the form of a
dipole pattern, but it varies very slowly with the angle $\Omega $ compared
with other contributions in $f_{\Omega }\left( \Omega \right) $, so we drop
it off when considering the spatial structure of the signal mode. Under this
circumstance, $f_{\Omega }\left( \Omega \right) $ is simply expressed as%
\begin{eqnarray}
f_{\Omega }\left( \Omega \right) &=&\left\langle u_{\perp }\left( {\bf r}%
\right) e^{-i\Delta {\bf k\cdot r}_{i}}\right\rangle _{\left\{ {\bf r}%
_{i}\right\} }  \label{18} \\
&=&\int d^{3}{\bf r}u_{\perp }\left( {\bf r}\right) p_{\text{dis}}\left(
{\bf r}\right) e^{ik_{0}z\left( 1-\cos \theta \right) -ik_{0}\sin \theta
\left( x\cos \varphi +y\sin \varphi \right) }.  \nonumber
\end{eqnarray}%
To further simplify this expression, we assume a Gaussian pump beam with $%
u_{\perp }\left( {\bf r}\right) =e^{-\left( x^{2}+y^{2}\right) /r_{0}^{2}}$,
and a Gaussian form for the transverse part of the atomic distribution
function with \cite{18}
\begin{eqnarray}
p_{\text{dis}}\left( {\bf r}\right) &=&\frac{1}{\pi LR_{0}^{2}}e^{-\left(
x^{2}+y^{2}\right) /R_{0}^{2}},\text{ (}-L/2\leq z\leq L/2),  \nonumber \\
p_{\text{dis}}\left( {\bf r}\right) &=&0,\text{ \ \ \ \ \ \ \ \ \ \ \ \ \ }%
\left( z<-L/2,\text{ or, }z>L/2\right)  \label{19e}
\end{eqnarray}%
where $r_{0}$ and $R_{0}$ characterize the radius of the pump beam and the
radius of the atomic ensemble, respectively, and $L$ is the length of the
ensemble. In this case, we have an analytic expression for the signal mode
function
\begin{equation}
f_{\Omega }\left( \Omega \right) =\frac{r_{0}^{2}}{r_{0}^{2}+R_{0}^{2}}e^{-%
\frac{1}{4}k_{0}^{2}r_{0}^{2}R_{0}^{2}\sin ^{2}\theta /\left(
r_{0}^{2}+R_{0}^{2}\right) }%
\mathop{\rm sinc}%
\left( k_{0}L\sin ^{2}\frac{\theta }{2}\right) ,  \label{20e}
\end{equation}%
where the function $%
\mathop{\rm sinc}%
$ is defined as $%
\mathop{\rm sinc}%
\left( x\right) \equiv \sin \left( x\right) /x$. From this expression, we
see that the signal photon mainly goes to the forward direction. The signal
mode is inside a small cone around $\theta =0$ with $\Delta \theta $
characterized by $\min \left[ \sqrt{r_{0}^{2}+R_{0}^{2}}/\left(
k_{0}r_{0}R_{0}\right) ,1/\sqrt{k_{0}L}\right] $.

We have shown that the first component of the solution (14) exactly
contributes to the ideal coherent process, and have specified the spatial
structure of the signal mode. Now we analyze the contributions of the noise
described by the second component of Eq. (14). From Eq. (14), we see that
the noise component $p_{2}\rho _{n}$ is expressed as a tensor product of the
atomic density operator $\rho _{n}^{a}=\sum_{i=1}^{N_{a}}|s\rangle
_{i}\left\langle s\right| $ and the remaining optical density operator $\rho
_{n}^{p}$. Besides the symmetric collective atomic mode $s_{s}$ and the
signal optical mode $a_{s}$, there are many other noise modes contributing
to the atomic density operator $\rho _{n}^{a}$ and the optical density
operator $\rho _{n}^{p}$. These noise modes correlate with each other in a
complicated way, however, these correlations do not contribute to the noise
of the relevant dynamics if the two signal modes are not involved in the
correlations, since the reduced density operator of the modes $s_{s}$ and $%
a_{s}$, which describes the relevant dynamics, will not be influenced by
these correlations after we take trace over all the noise modes.
Nevertheless, the correlations between the good modes and the noisy mode
will have influence on the relevant dynamics, and these contribute to the
two sources of noise we have mentioned before: the spontaneous emission loss
and the inherent mode-matching inefficiency. These two sources of noise are
described quantitatively by the conditional probabilities $p_{\text{spon}}$
and $p_{\text{mode}}$, respectively. The $p_{\text{spon}}$ represents the
probability that an atom is excited to the right mode $s_{s}$, but the
accompanying photon does not go to the signal mode $a_{s}$. From the
solution (14) to the whole density operator $\rho _{f}$, this possibility
can be expressed as
\begin{equation}
p_{\text{spon}}=1-\frac{tr\left( \left\langle 0_{a}0_{p}\right|
a_{s}s_{s}\rho _{f}s_{s}^{\dagger }a_{s}^{\dagger }|0_{a}0_{p}\rangle
\right) }{tr\left( \left\langle 0_{a}\right| s_{s}\rho _{f}s_{s}^{\dagger
}|0_{a}\rangle \right) },  \label{21e}
\end{equation}%
where $tr\left( \cdots \right) $ represents the trace over all the remaining
atomic and optical modes involved in the operator $\rho _{f}$. Similarly, $%
p_{\text{mode}}$ represents the probability that a photon is emitted to the
signal mode $a_{s}$, but the accompanying atomic excitation is not in the
right mode $s_{s}$, and this possibility can be expressed from the density
operator $\rho _{f}$ as%
\begin{equation}
p_{\text{mode}}=1-\frac{tr\left( \left\langle 0_{a}0_{p}\right|
a_{s}s_{s}\rho _{f}s_{s}^{\dagger }a_{s}^{\dagger }|0_{a}0_{p}\rangle
\right) }{tr\left( \left\langle 0_{p}\right| a_{s}\rho _{f}a_{s}^{\dagger
}|0_{p}\rangle \right) }.  \label{22e}
\end{equation}%
In the following, we need to calculate these two probabilities to quantify
the noise magnitudes.

From the solution (14), we can derive the probability $tr\left( \left\langle
0_{a}\right| s_{s}\rho _{f}s_{s}^{\dagger }|0_{a}\rangle \right) $ for one
atomic excitation in the mode $s_{s}$, the possibility $tr\left(
\left\langle 0_{p}\right| a_{s}\rho _{f}a_{s}^{\dagger }|0_{p}\rangle
\right) $ for one photon in the mode $a_{s}$, and the joint possibility $%
tr\left( \left\langle 0_{a}0_{p}\right| a_{s}s_{s}\rho _{f}s_{s}^{\dagger
}a_{s}^{\dagger }|0_{a}0_{p}\rangle \right) $ for one atom in the mode $%
s_{s} $ and one photon in the mode $a_{s}$. They are respectively given by
\begin{equation}
tr\left( \left\langle 0_{a}\right| s_{s}\rho _{f}s_{s}^{\dagger
}|0_{a}\rangle \right) =p_{c}+p_{2}/N_{a},  \label{23e}
\end{equation}

\begin{equation}
tr\left( \left\langle 0_{p}\right| a_{s}\rho _{f}a_{s}^{\dagger
}|0_{p}\rangle \right) =p_{c}+\chi p_{c},  \label{24e}
\end{equation}%
\begin{equation}
tr\left( \left\langle 0_{a}0_{p}\right| a_{s}s_{s}\rho _{f}s_{s}^{\dagger
}a_{s}^{\dagger }|0_{a}0_{p}\rangle \right) =p_{c}+\chi p_{c}/N_{a},
\label{25e}
\end{equation}%
where $p_{2}$ and $p_{c}$ are obtained respectively from the normalization
of $\rho _{n}$ and $a_{s}^{\dagger }$ ( see Eq. (16) and Eq. (17)), and the
ratio $\chi $ is defined as%
\begin{equation}
\chi =N_{a}tr\left( \left\langle 0_{p}\right| a_{s}\rho
_{n}^{p}a_{s}^{\dagger }|0_{p}\rangle \right) /p_{c},  \label{26e}
\end{equation}%
with $\rho _{n}^{p}$, the optical part of the noise component $p_{2}\rho
_{n} $ as we have specified before.

From the above equations, we see that we only need to calculate the two
ratios $p_{c}/p_{2}$ and $\chi $ to determine the noise probabilities $p_{%
\text{spon}}$ and $p_{\text{mode}}$. As we have mentioned before, the shape $%
u\left( {\bf r},t\right) $ of the pump beam is typically decomposed as $%
u\left( {\bf r},t\right) =u_{\perp }\left( {\bf r}\right) \mu _{l}\left(
z,t\right) $ with $\mu _{l}\left( z,t\right) $ approximately independent of $%
z$ along the size of the atomic ensemble. In this case, the two ratios $%
p_{c}/p_{2}$ and $\chi $ have much simplified expressions, which, become
independent of the interaction details, such as the interaction time or the
bandwidth of the coupling field, and depend only on some spatial
integrations determined by the geometry of the atomic ensemble and the pump
beam. For simplicity, we also neglect the slow variation of the coupling
coefficient $g_{{\bf k}}$ with the direction of ${\bf k}$. Under these
conditions, the ratios $p_{c}/p_{2}$ and $\chi $ are given respectively by
\begin{equation}
\frac{p_{c}}{p_{2}}=\frac{\int_{4\pi }d\Omega \left\{ \left| \left\langle
u_{\perp }\left( {\bf r}_{i}\right) e^{-i\Delta {\bf k\cdot r}%
_{i}}\right\rangle _{\left\{ {\bf r}_{i}\right\} }\right| ^{2}\right\} }{%
\int_{4\pi }d\Omega \left\{ \left\langle \left| u_{\perp }\left( {\bf r}%
_{i}\right) \right| ^{2}\right\rangle _{\left\{ {\bf r}_{i}\right\} }-\left|
\left\langle u_{\perp }\left( {\bf r}_{i}\right) e^{-i\Delta {\bf k\cdot r}%
_{i}}\right\rangle _{\left\{ {\bf r}_{i}\right\} }\right| ^{2}\right\} },
\label{27}
\end{equation}%
\begin{eqnarray}
\chi &=&\left[ \int_{4\pi }d\Omega \left\{ \left| \left\langle u_{\perp
}\left( {\bf r}_{i}\right) e^{-i\Delta {\bf k\cdot r}_{i}}\right\rangle
_{\left\{ {\bf r}_{i}\right\} }\right| ^{2}\right\} \right] ^{-2}  \nonumber
\\
&&\times \int_{4\pi }d\Omega \int_{4\pi }d\Omega ^{\prime }\left\{
\left\langle \left| u_{\perp }\left( {\bf r}_{i}\right) \right|
^{2}e^{-i\left( {\bf k-k}^{\prime }\right) {\bf \cdot r}_{i}}\right\rangle
_{\left\{ {\bf r}_{i}\right\} }\right.  \label{28e} \\
&&\times \left. \left\langle u_{\perp }^{\ast }\left( {\bf r}_{i}\right)
e^{i\Delta {\bf k\cdot r}_{i}}\right\rangle _{\left\{ {\bf r}_{i}\right\}
}\left\langle u_{\perp }\left( {\bf r}_{i}\right) e^{-i\Delta {\bf k}%
^{\prime }{\bf \cdot r}_{i}}\right\rangle _{\left\{ {\bf r}_{i}\right\}
}\right\} -1.  \nonumber
\end{eqnarray}%
Similar to the case for calculating the signal mode structure $f_{\Omega
}\left( \Omega \right) $, we also assume a Gaussian pump beam and a Gaussian
form for the transverse atomic distribution function as is shown by Eq.
(19). In this case, we have the following analytic expressions for these two
ratios
\begin{eqnarray}
\frac{p_{c}}{p_{2}} &\simeq &\frac{r_{0}^{2}\left(
r_{0}^{2}+2R_{0}^{2}\right) }{2\left( r_{0}^{2}+R_{0}^{2}\right) ^{2}}%
\int_{0}^{\pi }\sin \theta d\theta \left\{ e^{-\frac{1}{2}%
k_{0}^{2}r_{0}^{2}R_{0}^{2}\sin ^{2}\theta /\left(
r_{0}^{2}+R_{0}^{2}\right) }\right.  \nonumber \\
&&\left. \times
\mathop{\rm sinc}%
^{2}\left( k_{0}L\sin ^{2}\frac{\theta }{2}\right) \right\} ,  \label{29e}
\end{eqnarray}%
\begin{eqnarray}
\chi &=&\frac{\left( r_{0}^{2}+R_{0}^{2}\right) ^{2}}{r_{0}^{2}\left(
r_{0}^{2}+2R_{0}^{2}\right) }\left[ \int_{0}^{\pi }\sin \theta d\theta
\left\{ e^{-\frac{1}{2}k_{0}^{2}r_{0}^{2}R_{0}^{2}\sin ^{2}\theta /\left(
r_{0}^{2}+R_{0}^{2}\right) }%
\mathop{\rm sinc}%
^{2}\left( k_{0}L\sin ^{2}\frac{\theta }{2}\right) \right\} \right] ^{-2}
\nonumber \\
&&\times \int_{0}^{\pi }\int_{0}^{\pi }\sin \theta \sin \theta ^{\prime
}d\theta d\theta ^{\prime }\left\{ \frac{1}{2\pi }\int_{0}^{2\pi }e^{-\frac{1%
}{4}k_{0}^{2}r_{0}^{2}R_{0}^{2}\left( \sin ^{2}\theta +\sin ^{2}\theta
^{\prime }-2\sin \theta \sin \theta ^{\prime }\cos \varphi \right) /\left(
r_{0}^{2}+2R_{0}^{2}\right) }d\varphi \right.  \label{30} \\
&&\left. \times e^{-\frac{1}{4}k_{0}^{2}r_{0}^{2}R_{0}^{2}\left( \sin
^{2}\theta +\sin ^{2}\theta ^{\prime }\right) \left(
r_{0}^{2}+R_{0}^{2}\right) }%
\mathop{\rm sinc}%
\left[ k_{0}L\left( \sin ^{2}\frac{\theta }{2}-\sin ^{2}\frac{\theta }{2}%
\right) \right]
\mathop{\rm sinc}%
\left( k_{0}L\sin ^{2}\frac{\theta }{2}\right)
\mathop{\rm sinc}%
\left( k_{0}L\sin ^{2}\frac{\theta ^{\prime }}{2}\right) \right\} -1.
\nonumber
\end{eqnarray}%
In writing Eq. (29), we have assumed $k_{0}L\gg 1$ and have neglected the
term which is about $1/k_{0}L$ times smaller.

From the two ratios $p_{c}/p_{2}$ and $\chi $, the spontaneous emission loss
and the inherent mode matching inefficiency are directly written as%
\begin{equation}
p_{\text{spon}}\simeq \left( 1+N_{a}p_{c}/p_{2}\right) ^{-1},  \label{31}
\end{equation}%
\begin{equation}
p_{\text{mode}}\simeq \chi /\left( 1+\chi \right) .  \label{32}
\end{equation}%
The approximations in Eqs. (31) and (32) are valid under the condition $%
N_{a}\gg 1$\ (we neglected the terms which are $1/N_{a}$ times smaller). To
minimize the two sources of noise $p_{\text{spon}}$ and $p_{\text{mode}}$,
it is better to have a large $N_{a}p_{c}/p_{2}$ and a small $\chi $. We will
discuss the details in the following section on how to minimize these two
noise by using different kinds of mode matching methods.

At the end of this section, we would like to make a brief comparison with
the single atom case. For the case of a single atom, if the position of this
atom is also fluctuating in the space, the above calculation is still valid
but with $N_{a}=1$. From Eqs. (21)-(26), we see that in the case of $N_{a}=1$%
, we always have $p_{\text{mode}}=0$. Thus, for the case of a single atom,
the spontaneous emission loss is the only source of noise, as has been shown
intuitively in Fig. 4. However, in this case, the spontaneous emission
inefficiency $p_{\text{spon}}$, which can be approximated by $p_{\text{spon}%
}\simeq 1-\left( 1+\chi \right) p_{c}/p_{2}$ since $p_{c}\ll p_{2}$, becomes
much larger. If we compare the efficiency (defined as $1-p_{\text{spon}}$)
from the spontaneous emission noise between the single-atom case and the
atomic-ensemble case, we see that this efficiency is increased by about a
factor of $N_{a}$ for the atomic ensemble. This is what we called the
collective enhancement of the signal-to-noise ratio for the many-atom
ensemble.

\section{Mode matching methods}

The first mode-matching method is to choose the signal mode as the mode for
detection, with the mode function exactly given by the inherent spatial
structure (20) of the signal light. The mode function (20) is similar to a
Gaussian function, especially when $k_{0}r_{0}^{2}>L$ and $k_{0}R_{0}^{2}>L$%
. (It would be an exact Gaussian function if the axial distribution of $p_{%
\text{dis}}\left( {\bf r}\right) $ is Gaussian). One can couple a Gaussian
mode into a single-mode optical fibre with a good technical mode-matching
efficiency. For this case with an exact mode matching, the two noise
probabilities $p_{\text{spon}}$ and $p_{\text{mode}}$ are exactly given by
Eqs. (29)-(32). We now go to the calculations of these two noise
probabilities. For this purpose, we need to understand first how to get a
large $N_{a}p_{c}/p_{2}$ and a small $\chi $ to minimize the noise $p_{\text{%
spon}}$ and $p_{\text{mode}}$.

First, we give an estimation of the ratio $N_{a}p_{c}/p_{2}$. It has a
definite physical meaning. From Eq. (29), we see that the integration
function is significantly different to zero only when the integration
variable $\theta $ falls into a small cone with $0\leq \theta <\theta _{f}$,
where $\theta _{f}$ is estimated by $\theta _{f}\sim \min \left( 1/\sqrt{%
k_{0}L},1/k_{0}r_{eff}\right) $ with $r_{eff}\equiv r_{0}R_{0}/\sqrt{%
r_{0}^{2}+R_{0}^{2}}$. Thus, the result of the integration can be estimated
by $\theta _{f}^{2}$. We also know that the total atom number can be
approximated by $N_{a}\sim n_{a}\pi R_{0}^{2}L$, where $n_{a}$ denotes the
average atom number density. From this, we see that the ratio $%
N_{a}p_{c}/p_{2}$ is estimated by
\begin{equation}
\frac{N_{a}p_{c}}{p_{2}}\sim \min \left( \frac{1}{2}n_{a}\lambda
_{0}r_{eff}^{2},\text{ }\frac{1}{4\pi }n_{a}\lambda _{0}^{2}L\right) .
\label{33}
\end{equation}%
If we assume that the two bounds for $N_{a}p_{c}/p_{2}$ are comparable,
which means that $1/\sqrt{k_{0}L}\sim 1/k_{0}r_{eff}$, or the Fresnel number
defined as $Fr\equiv \pi r_{eff}^{2}/\lambda _{0}L$ is in the order of $1$,
we have $N_{a}p_{c}/p_{2}\sim n_{a}\lambda _{0}^{2}L/\left( 4\pi \right)
\sim d_{o}/\left( 4\pi \right) $, where the on-resonance optical depth $%
d_{o} $ is simply defined as $d_{o}\equiv n_{a}\lambda _{0}^{2}L$.
Therefore, to have a small spontaneous emission inefficiency $p_{\text{spon}}
$, we need to use an optically dense ensemble with $d_{o}\gg 1$ (the
off-resonance optical depth can still be much smaller than $1$ due to the
large detuning).

Next, we consider how to minimize the ratio $\chi $ by choosing appropriate
interaction configurations. From Eq. (30), we see that $\chi $ increase with
the ratio $R_{0}/r_{0}$. This can be intuitively understood as follows: the
noise characterized by $p_{\text{mode}}$ and $\chi $ comes from the density
fluctuation of the atomic ensemble. With a larger $R_{0}$, the atoms have
more space to move around, and one thus has a relatively larger density
function and a larger noise ratio $\chi $ \cite{19}. Thus, to minimize $\chi
$, we choose a configuration with $r_{0}\gg R_{0}$ in the following, which
means, the pumping laser is shined on all the atoms with a broad cross
section. In this case, the effective radius $r_{eff}\simeq R_{0}$.

Now we would like to calculate the noise probabilities $p_{\text{spon}}$ and
$p_{\text{mode}}$ in a more accurate way in the case of exact mode matching.
This requires us to carry out the complicate integrations in Eqs. (29) and
(30), which is only possible by using numerical methods. The spontaneous
emission inefficiency $p_{\text{spon}}$ depends on the optical depth $%
d_{o}=n_{a}\lambda _{0}^{2}L$. We have numerically calculated the value of\ $%
p_{\text{spon}}$ versus the optical depth $d_{o}$ under different geometries
of the atomic ensemble. The geometry of the ensemble is described by the
Fresnel number $Fr\equiv \pi R_{0}^{2}/\lambda _{0}L$. The calculation
result is shown in Fig. 5. From the figure, we see that the spontaneous
emission inefficiency is insensitive to the geometry of the ensemble. The
two curves with $Fr=1$ and $Fr=10$ basically overlaps. With a much smaller $%
Fr=0.1$, the inefficiency $p_{\text{spon}}$ increases by about a factor of $%
2 $. On the other hand, $p_{\text{spon}}$ is sensitive to the optical depth $%
d_{o}$. It decrease with $1/d_{o}$ approximately linearly, which confirms
the rough estimation from the above. From the curves, we can write $p_{\text{%
spon}}$ approximately as $p_{\text{spon}}=1/(1+d_{o}/26)$ for the Fresnel
number $F_{r}$ from $1$ to $10$. In the above numerical calculation, we
assumed some typical values for the parameters with $\lambda _{0}\sim 0.8\mu
$m and $L\sim 1$cm, but actually the result is very insensitive to the
parameter $k_{0}L$ as long as it is still much larger than $1$. For an
atomic cell with a typical number density $n_{a}\sim 10^{12}/$cm$^{3}$ and
length $L\sim 1$cm, the optical depth $d_{o}\sim 6.4\times 10^{3}$, and we
have a spontaneous emission inefficiency $p_{\text{spon}}\sim 0.4\%$, which
is very small.
\begin{figure}[tbp]
\epsfig{file=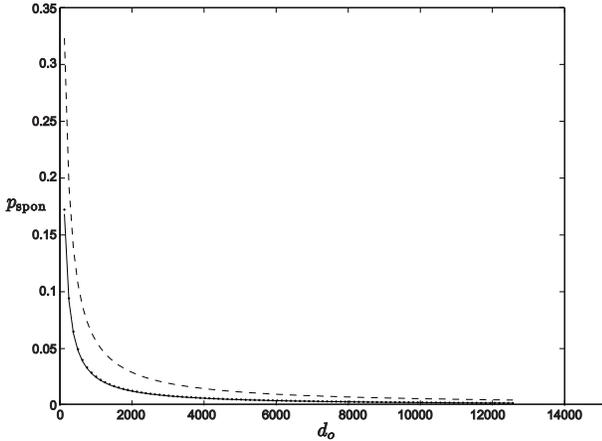,width=8cm} \caption{The spontaneous
emission inefficiency $p_{\text{spon}}$ versus the on-resonance
optical depth $d_{o}$ of the ensemble under different geometries
characterized by the Fresnel number $Fr$. We have $Fr=10$ for the
solid curve, $Fr=1$ for the dotted curve, and $Fr=0.1$ for the
dashed curve.}
\end{figure}

The inherent mode matching inefficiency $p_{\text{mode}}$ is determined by
the geometry of the ensemble, and does not depend on the value of the
optical depth. We have numerically calculated the value of $p_{\text{mode}}$%
. It turns out that $p_{\text{mode}}$\ is actually also insensitive to the
geometric parameters $k_{0}L$ and $Fr$ as long as they are still in the
typical region. For instance, we have $p_{\text{mode}}\simeq 24\%$ for $Fr=1$%
, $p_{\text{mode}}\simeq 25\%$ for $Fr=10$, and $p_{\text{mode}}\simeq 23\%$
for $Fr=0.1$. In each case, we change the length $L$ from $0.3$cm to $3$cm\
(with $\lambda _{0}\sim 0.8\mu $m), and the value of $p_{\text{mode}}$
basically does not change at all. Note that the typical inherent mode
matching inefficiency $p_{\text{mode}}\simeq 24\%$ is quite large, and we
can not efficiently reduce it by controlling the geometry of the ensemble.
As we have mentioned before, the mode mismatching noise has a more
significant influence on the efficiency of the application schemes than the
spontaneous emission noise. The large inherent mode matching inefficiency is
a disadvantage of the exact mode-matching method.

To reduce the inefficiency $p_{\text{mode}}$, we consider the following
simple improvement to the exact mode matching method. After the
light-atomic-ensemble interaction, the scattered light is focused by a lens,
and we apply an aperture on the focal plane of the lens to collect the
signal light only from a small cone in the forward direction with $0\leq
\theta \leq \theta _{D}$, where the detection angle $\theta _{D}$ can be
controlled by the size of the aperture. We call this kind of mode matching
the filtered exact mode matching. By the exact mode matching, we detect the
photon scattered to the signal mode; and by the filtered exact mode
matching, we only detect the scattered photon which is in the signal mode
and at the same time lies in the filtering cone with $0\leq \theta \leq
\theta _{D}$. Due to the additional restriction, we now have a less success
probability to register the photon. The spontaneous emission inefficiency $%
p_{\text{spon}}$ should increase, but through this scarification, we can
significantly reduce the noise $p_{\text{mode}}$, which means, if we
register the photon with this additional restriction, we are more confident
that the accompanying atomic excitation will go to the collective atomic
mode $s_{s}$. To calculate $p_{\text{spon}}$ and $p_{\text{mode}}$\ for the
filtered exact mode matching, we can still use Eqs. (21)-(26), but the
signal mode $a_{s}$ should be confined inside the small cone with $0\leq
\theta \leq \theta _{D}$. Inside this filtering cone, the signal mode has
the same mode structure as given by Eq. (20). With this modified signal
mode, we find that $p_{\text{mode}}$ is still given by Eqs. (32) and (30),
but all the integrations of $\theta $ and $\theta ^{\prime }$ in Eq. (30)
can only be taken from $0$ to the filtering angle $\theta _{D}$ instead of
from $0$ to $\pi $. The spontaneous emission inefficiency $p_{\text{spon}}$
now has the following expression%
\begin{equation}
p_{\text{spon}}\simeq 1-\frac{N_{a}p_{c}|_{\theta \leq \theta _{D}}}{%
N_{a}p_{c}+p_{2}},  \label{34}
\end{equation}%
where we have assumed $N_{a}\gg 1$ for the approximation, and the symbol $%
p_{c}|_{\theta \leq \theta _{D}}$ means that the integration of $\theta $ in
$p_{c}$ can only be taken from $0$ to $\theta _{D}$. From Eq. (34), we need
to calculate two ratios $p_{c}/p_{2}$ and $p_{c}|_{\theta \leq \theta
_{D}}/p_{2}$ for $p_{\text{spon}}$. The ratio $p_{c}/p_{2}$ is exactly given
by Eq. (29), and the ratio $p_{c}|_{\theta \leq \theta _{D}}/p_{2}$ has the
same form as Eq. (29), but the integration of $\theta $ is only from $0$ to $%
\theta _{D}$.

We have numerically calculate the noise probabilities $p_{\text{spon}}$ and $%
p_{\text{mode}}$ versus the filtering angle $\theta _{D}$, and the result is
shown in Fig. 6 with the Fresnel number $Fr=1$ and the optical depth $%
d_{o}\sim 1.9\times 10^{3}$. From the figure, we see that by decreasing the
filtering angle $\theta _{D}$, we can significantly reduce the noise $p_{%
\text{mode}}$. The cost is that at the same time the inefficiency $p_{\text{%
spon}}$ significantly increases. However, as we have mentioned before,
though both of the noise $p_{\text{spon}}$ and $p_{\text{mode}}$ are
correctable in the application schemes in Refs. \cite{12,13}, the mode
mismatching noise $p_{\text{mode}}$ has a more severe influence on the final
efficiency of the schemes than the spontaneous emission noise $p_{\text{spon}%
}$. Thus, for these applications, it is worthy to choose an appropriate
filtering angle $\theta _{D}$ with $p_{\text{mode}}$ significantly smaller
than $p_{\text{spon}}$. For instance, if we choose the angle $\theta
_{D}\simeq 0.002$, the mode mismatching noise $p_{\text{mode}}\simeq 0.9\%$,
which is basically negligible compared with other sources of noise. At the
same time, $p_{\text{spon}}\simeq 66.6\%$, which seems to be quite large.
However, to overcome this noise, we need only to increase the repetitions of
the entanglement generation scheme in Refs. \cite{12,13} by another factor
of $3$ (given by $1/\left( 1-p_{\text{spon}}\right) $), which is actually
only a moderate cost. This choice would be much better than the case without
any filtering, where we have an inherent mode mismatching noise $p_{\text{%
mode}}\simeq 24\%$. For the purpose of guiding the choice of the best
experimental configuration, we list several important values of $p_{\text{%
spon}}$ and $p_{\text{mode}}$ versus the angle $\theta _{D}$ in the
following table

\begin{table}[tbp]
\begin{tabular}{llllll}
$\theta _{D}$ & $0.0015$ & $0.0020$ & $0.0025$ & $0.0040$ & $0.0055$ \\
$p_{\text{mode}}$ & $0.31\%$ & $0.92\%$ & $2.05\%$ & $8.98\%$ & $17.5\%$ \\
$p_{\text{spon}}$ & $79.6\%$ & $66.6\%$ & $52.9\%$ & $19.5\%$ & $5.02\%$%
\end{tabular}%
\end{table}

\begin{figure}[tbp]
\epsfig{file=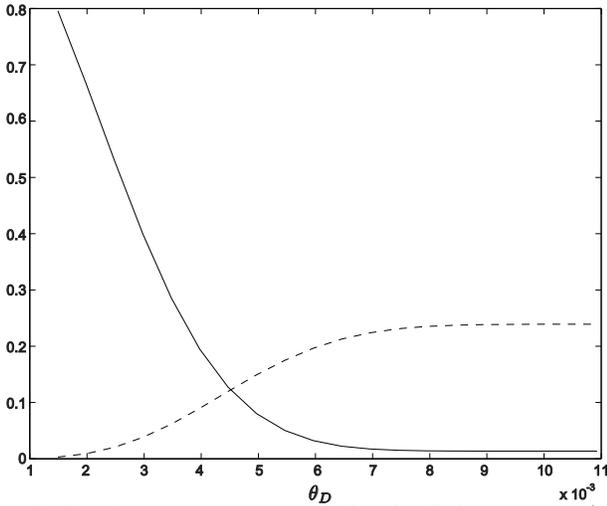,width=8cm} \caption{The spontaneous
emission inefficiency $p_{\text{spon}}$ (solid curve) and the
inherent mode mismatching probability $p_{\text{mode}}$ (dashed
curve) versus the filtering angle $\protect\theta _{D}$ with the
Fresnel number $Fr=1$.}
\end{figure}

We have also calculated the noise probabilities $p_{\text{spon}}$ and $p_{%
\text{mode}}$ versus the angle $\theta _{D}$ under different geometries of
the ensemble characterized by the Fresnel number $Fr$. For $Fr=10$ and $%
Fr=0.1$, the results are shown in Fig. 7. In all the calculations, we
assumed the same optical depth $d_{o}\sim 1.9\times 10^{3}$ and the cell
length $L\sim 1$cm. The qualitative properties of the curves for different
Fresnel numbers are quite similar, but one needs to shift the appropriate
angle $\theta _{D}$. With a large Fresnel number $Fr$, one needs to choose a
smaller angle $\theta _{D}$. The appropriate filtering angle $\theta _{D}$
changes by a factor of $2$-$3$ if the Fresnel number changes by a factor of $%
10$.
\begin{figure}[tbp]
\epsfig{file=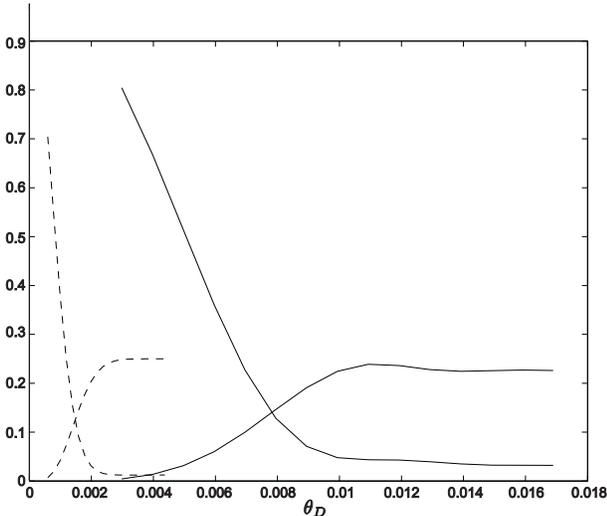,width=8cm} \caption{The spontaneous
emission inefficiency $p_{\text{spon}}$ and the inherent mode
mismatching probability $p_{\text{mode}}$ versus the filtering
angle $\protect\theta _{D}$ with the Fresnel number $Fr=10$ (the
two dashed curves) and $Fr=0.1$ (the two solid curves). The
spontaneous emission inefficiency increases with the decrease of
the angle $\protect\theta _{D}$ while the mode mismatching noise
decreases.}
\end{figure}

Finally, we describe a mode matching method which is simplest for
implementation. For this method, we still use an aperture to select the
scattered light from the small cone with $0\leq \theta \leq \theta _{D}$.
But now we do not use any technical mode matching to choose only the signal
mode with the right mode structure $f_{\Omega }\left( \Omega \right) $ for
detection. Instead, we detect the photon in all the modes which lie in the
small cone with $0\leq \theta \leq \theta _{D}$. We call this the simple
filtering method. This seems to be a bad mode-matching method, since one
does not make any distinction between the modes in this small cone. If the
registered photon does not come from the right mode $a_{s}$, the
accompanying atomic excitation will be most probably not in the collective
mode $s_{s}$, and one thus has a significantly larger inherent mode
mismatching noise $p_{\text{mode}}$. However, the observation here is that
if $\theta _{D}$ is sufficiently small, basically only the signal mode
exists in this small cone, and all the other modes make negligible
contributions. The important question is how small the angle $\theta _{D}$
should be to guarantee a small $p_{\text{mode}}$. Note that for the simple
filtering method, the spontaneous emission inefficiency $p_{\text{spon}}$
can be calculated in the same way as the filtered exact mode matching method
by the use of Eq. (34). But we have a different expression for $p_{\text{mode%
}}$, which in this case is defined as the probability that the accompanying
atomic excitation does not go to the collective mode $s_{s}$ when one photon
(from any optical modes) is detected in the small cone with $0\leq \theta
\leq \theta _{D}$. This possibility can be expressed as
\begin{equation}
p_{\text{mode}}\simeq 1-\frac{p_{c}|_{\theta \leq \theta _{D}}+\left(
1/N_{a}\right) tr\left( p_{2}\rho _{n}\right) |_{\theta \leq \theta _{D}}}{%
p_{c}|_{\theta \leq \theta _{D}}+tr\left( p_{2}\rho _{n}\right) |_{\theta
\leq \theta _{D}}},  \label{35}
\end{equation}%
where the trace $tr\left( \cdots \right) $ is over all the atomic and the
optical modes, and the symbol $|_{\theta \leq \theta _{D}}$ has the same
meaning that the integrations of $\theta $ in $p_{c}$ and $tr\left(
p_{2}\rho _{n}\right) $ are only from $0$ to $\theta _{D}$. In the limit of $%
N_{a}\gg 1$, $p_{\text{mode}}$ can still be written into the form $p_{\text{%
mode}}\simeq \chi _{s}/\left( 1+\chi _{s}\right) $, with
\begin{eqnarray}
\frac{1}{\chi _{s}+1} &=&\frac{p_{c}|_{\theta \leq \theta _{D}}}{tr\left(
p_{2}\rho _{n}\right) |_{\theta \leq \theta _{D}}+p_{c}|_{\theta \leq \theta
_{D}}}  \nonumber \\
&=&\frac{1}{\left( 1-\cos \theta _{D}\right) }\int_{0}^{\theta _{D}}\sin
\theta d\theta \left\{ e^{-\frac{1}{2}k_{0}^{2}R_{0}^{2}\sin ^{2}\theta
}\right.  \nonumber \\
&&\left. \times
\mathop{\rm sinc}%
^{2}\left( k_{0}L\sin ^{2}\frac{\theta }{2}\right) \right\} .  \label{36e}
\end{eqnarray}%
In writing Eq. (36), we have assumed the same kind of atomic distribution
function $p_{\text{dis}}\left( {\bf r}\right) $ and the pump mode function $%
u_{\perp }\left( {\bf r}\right) $ as we have specified before. We can use
Eqs. (36) to numerically calculate the mode mismatching probability $p_{%
\text{mode}}$ versus the filtering angle $\theta _{D}$ for the simple
filtering method. The result, together with the curve for $p_{\text{spon}}$
versus $\theta _{D}$, is shown in Fig. 8 for the Fresnel number $Fr=1$. In
the calculation, we assumed the same optical depth $d_{o}\sim 1.9\times
10^{3}$. From the figure, we see that if $\theta _{D}$ is large, the mode
mismatching probability $p_{\text{mode}}$ is very large, but as $\theta _{D}$
decreases, $p_{\text{mode}}$ can still tend to zero. This confirms our
intuitive observation. With the same parameters, for the simple filtering
method, we should further decrease the filtering angle $\theta _{D}$ by a
factor of $2$-$3$ to get the optimal configuration compared with the
filtered exact mode matching method. Of course, due to the decrease of the
optimal $\theta _{D}$, the corresponding spontaneous emission inefficiency $%
p_{\text{spon}}$ significantly increases. To guide the experimental choice
of the optimal $\theta _{D}$, we also list some important values of $p_{%
\text{spon}}$ and $p_{\text{mode}}$ for the simple filtering method

\begin{table}[tbp]
\begin{tabular}{llllll}
$\theta _{D}$ & $0.0006$ & $0.0010$ & $0.0014$ & $0.0020$ & $0.0032$ \\
$p_{\text{mode}}$ & $1.76\%$ & $4.78\%$ & $9.10\%$ & $17.5\%$ & $37.4\%$ \\
$p_{\text{spon}}$ & $96.4\%$ & $90.3\%$ & $82.0\%$ & $66.6\%$ & $35.1\%$%
\end{tabular}%
\end{table}

\begin{figure}[tbp]
\epsfig{file=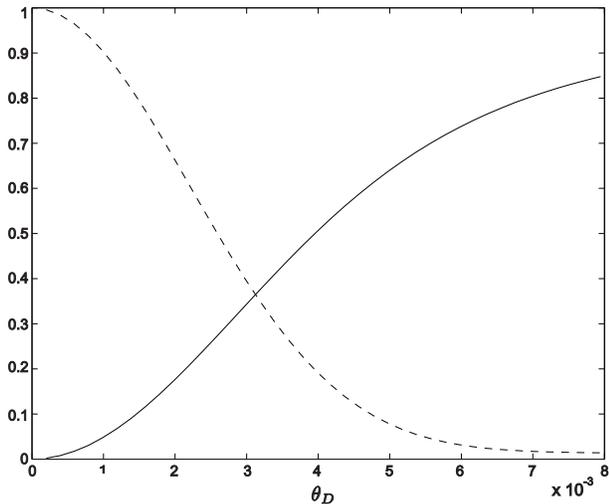,width=8cm} \caption{The spontaneous
emission inefficiency $p_{\text{spon}}$ (dashed curve) and the
inherent mode mismatching probability $p_{\text{mode}}$ (solid
curve) versus the filtering angle $\protect\theta _{D}$ with the
Fresnel number $Fr=1$ for the simple filtering method.}
\end{figure}

For the simple filtering method, we also calculated the noise probabilities $%
p_{\text{spon}}$ and $p_{\text{mode}}$ under different geometries of the
ensemble. For the Fresnel number $Fr=10$ and $Fr=0.1$, the results are shown
in Fig. 9. We assumed the same optical depth and the cell length in these
calculations. Since the qualitative picture revealed by this figure is quite
similar to the case of the filtered exact mode matching, we do not need to
discuss its properties in details.
\begin{figure}[tbp]
\epsfig{file=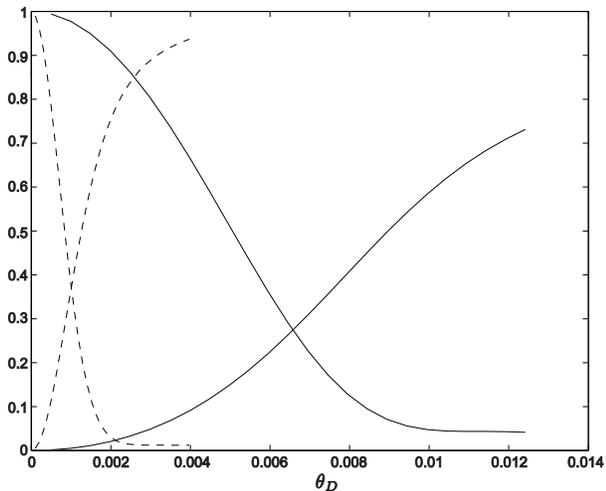,width=8cm} \caption{The spontaneous
emission inefficiency $p_{\text{spon}}$ and the inherent mode
mismatching probability $p_{\text{mode}}$ versus the filtering
angle $\protect\theta _{D}$ with the Fresnel number $Fr=10$ (the
two dashed curves) and $Fr=0.1$ (the two solid curves) for the
simple filtering method.}
\end{figure}

For experimental realizations, the simple filtering method can be much
easier than other mode matching methods. The calculation here shows that the
price we need to pay for this simplification is to choose a smaller
filtering angle with a significantly increased spontaneous emission
inefficiency. Note that for the application schemes in Refs. \cite{12,13},
this price seems to be still acceptable. For instance, if we choose $\theta
_{D}\simeq 0.001$, the mode mismatching noise $p_{\text{mode}}\simeq 4.8\%$
has already been small. In this case, the additional repetitions of the
entanglement generation scheme in Refs. \cite{12,13} is given by the factor $%
1/\left( 1-p_{\text{spon}}\right) \simeq 10$, which is not too much.

\section{Summary}

In summary, we have developed a theory in the weak pumping limit to describe
the three-dimensional effects in the interaction between the free-space
light and the many-atom ensemble. The calculations demonstrate some
interesting results which are not know from the other approaches: firstly,
it shows that the signal light has an inherent spatial mode structure, which
is determined together by the geometry of the ensemble and the mode
structure of the pump beam. Secondly, it reveals that there will be two
sources of noise during the light-atom interaction. One is the spontaneous
emission inefficiency, which is inversely proportional to the on-resonance
optical depth of the ensemble; and the other is the inherent mode
mismatching noise, which arises from the density fluctuation of the
ensemble, and should be fully determined by the geometry of the ensemble for
room-temperature atomic cells. It turns out that the inherent mode
mismatching noise is quite large if one collects the signal light through
the exact mode matching method, and one cannot efficiently reduce this noise
by optimizing the geometry of the ensemble since in the typical parameter
region, this noise is insensitive to the cell geometry. Finally, we show an
effective way to reduce the inherent mode mismatching noise by adding an
aperture to select the light only from a small emission cone. By this mean,
we can efficiently reduce the mode mismatching noise at the price of
increasing the spontaneous emission inefficiency. It is worthy to do this
since the spontaneous emission inefficiency is far less important for some
application schemes. There are two methods for this purpose: the filtered
exact mode matching method and the simple filtering method. Both methods can
reduce the inherent mode matching noise if one chooses an appropriate
filtering angle.

The calculations in this paper show that there is a large collective
enhancement of the signal-to-noise ratio if one compares the many-atom
ensemble with the single-atom case, which confirms the predication from some
simple theoretical models, and demonstrates the validity of this important
observation for the complicated realistic situations.

At the end of this paper, we should mention that the calculations here also
show that for any mode matching method, we can not make the spontaneous
emission inefficiency $p_{\text{spon}}$ and the inherent mode mismatching
noise $p_{\text{mode}}$ both negligible for the room-temperature atomic cell
even if the cell has a large on-resonance optical depth. This result does
not have much influence on the application schemes in Refs. \cite{12,13}
since they are inherently robust to the noise $p_{\text{spon}}$ and $p_{%
\text{mode}}$. As a consequence of this inherent robustness, a
non-negligible but not-huge noise only has a moderate influence on
the efficiency of the schemes. However, for any of the application
schemes of atomic ensembles which are not inherently robust, such
as the quantum light memory scheme \cite{7,8,,9} or the continuous
variable teleportation scheme \cite{4,5,6}, a non-negligible noise
will decrease the scheme fidelity. The calculations here do not
directly apply to these schemes, since they work out of the
perturbation region. However, it indeed raises the interesting
question whether there is some theoretical limit on the best
achievable fidelity of these schemes if one uses free-space atomic
cells for their implementation.

{\bf Acknowledgments}: L.M.D. thanks Jeff Kimble and Alex Kuzmich
for discussions. Work of L.M.D was supported by the Caltech MURI
Center for Quantum Networks under ARO Grant No. DAAD19-00-1-0374,
by the National Science Foundation under Grant No. EIA-0086038,
and also by the Chinese Science Foundation and Chinese Academy of
Sciences. Work at the University of Innsbruck was supported by the
Austrian Science Foundation and EU networks.


\begin{references}
\bibitem{17} Note that for rapidly moving atoms, we also have the effect of
Doppler shifts which are not included in the Hamiltonian (6). However, the
Doppler shifts are not important for the off-resonant coupling considered
here (they are still significantly than the detuning). In addition, the
Doppler shifts will cancel for the spontaneous emission modes which are
nearly collinear with the pumping laser (including the important signal
light mode).
\bibitem{18} The noise properties in \ subsequent discussions will depend on
the ratio of the transverse length scales for the atomic cell and for the
pump beam, but they are insensitive to the detailed form of the atomic
distribution function. Here, for simplicity of the calculation, we assume a
Gaussian form for the atomic transverse distribution function. But we expect
that all the results will be basically still valid if the transverse
distribution is not in the Gaussian form, but instead, is uniform within the
cell radius $R_{0}$ and zero outside.
\bibitem{19} Note that this is a valid picture for room-temperature atomic
cells. However, for ultracold atoms, each atom will be localized in a small
space, and the density fluctuation basically will not depend on the size and
geometry of the ensemble. Thus, for cold atoms, one may have a different
interaction picture and a smaller noise ratio.
\end{references}
\end{document}